\documentclass[12pt,a4paper]{article}
\usepackage{epsf}

\begin{document}

\author{R. Fiore\thanks{
\ fiore@fermi.fis.unical.it}, A. Tiesi\thanks{
\ tiesi@fermi.fis.unical.it} \\
Dipartimento di Fisica, Universit\`{a} della Calabria, \\
87036 Arcavacata di Rende, Cosenza, Italy \and L. Masperi\thanks{
\ masperi@cab.cnea.edu.ar}, A. M\'{e}gevand\thanks{
\ megevand@cab.cnea.edu.ar} \thanks{
\ Fellow of Consejo Nacional de Investigaciones Cient\'{\i}ficas y
T\'{e}cnicas, Argentina.} \\
Centro At\'{o}mico Bariloche and Instituto Balseiro, \\
Comisi\'{o}n Nacional de Energ\'{\i}a At\'{o}mica and Universidad Nacional
de Cuyo, \\
8400 San Carlos de Bariloche, Argentina}
\title{Effect of strong magnetic field on the first-order electroweak phase
transition}
\date{}
\maketitle

\begin{abstract}
The broken-symmetry electroweak vacuum is destabilized in the presence of a
magnetic field stronger than a critical value. Such magnetic field may be
generated in the phase transition and restore the symmetry inside the
bubbles. A numerical calculation indicates that the first-order phase
transition is delayed but may be completed for a sufficient low value of the
Higgs mass unless the magnetic field is extremely high.
\end{abstract}

\def\nullset{\hbox{\sf\Slo}}  
\def\dsty{\displaystyle}
\def\ssty{\scriptstyle}
\def\sssty{\scriptscriptstyle}

\def\gsim{\relax\leavevmode
     \ifmmode\mathchoice			
	{\raise1pt\hbox{$>$} \kern-0.65em \lower3.5pt\hbox{$\ssty \sim$}}
	{\raise1pt\hbox{$>$} \kern-0.65em \lower3.5pt\hbox{$\ssty \sim$}}
	{\raise0.75pt\hbox{$\ssty >$} \kern-0.56em
		\lower3.5pt\hbox{$\sssty \sim$}}
	{\raise0.85pt\hbox{$\sssty >$} \kern-0.50em
		\lower2.7pt\hbox{$\sssty \sim$}}
     \else				
	{$\raise1pt\hbox{$>$} \kern-0.65em
		\lower3.5pt\hbox{$\ssty \sim$}$}\fi}

\def\lsim{\relax\leavevmode
     \ifmmode\mathchoice			
	{\raise1pt\hbox{$<$} \kern-0.65em \lower3.5pt\hbox{$\ssty \sim$}}
	{\raise1pt\hbox{$<$} \kern-0.65em \lower3.5pt\hbox{$\ssty \sim$}}
	{\raise0.75pt\hbox{$\ssty <$} \kern-0.56em
		\lower3.5pt\hbox{$\sssty \sim$}}
	{\raise0.85pt\hbox{$\sssty <$} \kern-0.50em
		\lower2.7pt\hbox{$\sssty \sim$}}
     \else				
	{$\raise1pt\hbox{$<$} \kern-0.65em
		\lower3.5pt\hbox{$\ssty \sim$}$}\fi}

\section{Introduction}

It has been found that very strong magnetic fields are capable of
destabilizing the electroweak vacuum by forming a vector boson $W^{+}W^{-}$
condensate and restoring the symmetry \cite{amb}. The required field can
only be thought to have existed at the very beginning of the universe and
one of the possibilities is that it was generated during the electroweak
phase transition \cite{baym,cheng}. This primordial field may have been
subsequently the seed of the present galactic magnetic field \cite{enq}.

One may wonder whether the restoration of symmetry caused by this strong
magnetic field can delay the electroweak phase transition. In particular, if
it is of first-order the magnetic field might avoid its completion through
the bubble mechanism.

The simplest way to see why a strong magnetic field can destabilize the
electroweak vacuum is to consider the energy of a charged spin-one particle
interacting with a uniform magnetic field along the 3-axis
\begin{equation}
E_{N}^{2}=p_{3}^{2}+m_{0}^{2}+\left( 2N+1\right) eB-geB\quad .
\end{equation}

For the lowest Landau level $N=0$ if the gyromagnetic factor $g$ is $2$ as
occurs in the $W$ case, it is clear that the effective mass will become zero
for
\begin{equation}
B_{c}=\frac{m_{W}^{2}}{e}\simeq 10^{24}G\quad .  \label{eq2}
\end{equation}
This expression is analogous to that of the critical electric field required
to create pairs through tunneling.

If one wishes to calculate the decay probability of the vacuum, one must
evaluate
\begin{equation}
Z=<0|e^{-iHt}|0>=e^{-it\left( E_{vac}-i\frac{\Gamma }{2}\right) }\quad .
\end{equation}

In Euclidean metric the one-loop amplitude for a scalar field depends on $
\det \left( -D_{E}^{2}+m^{2}\right) $ with $D_{E\mu }=\partial _{\mu
}-ieA_{\mu }$, being $D_{E4}=iD_{0}$. Using the Schwinger proper time method
\cite{schw} one obtains
\begin{equation}
\ln Z=\int_{0}^{\infty }\frac{ds}{s}tre^{-\left( -D_{E}^{2}+m^{2}\right)
s}\quad .  \label{e2}
\end{equation}
For constant electromagnetic fields the trace is known to give the vacuum
energy density \cite{schm}
\begin{equation}
\rho =-\int_{0}^{\infty }\frac{ds}{s}\frac{e^{-m^{2}s}}{\left( 4\pi s\right)
^{2}}\left[ \frac{es\sqrt{E^{2}-B^{2}}}{\sin \left( es\sqrt{E^{2}-B^{2}}
\right) }-1\right] \quad ,  \label{e1}
\end{equation}
where the $-1$ comes from subtracting $\rho \left( A=0\right) $. This
integral has a logarithmic divergence for $s=0$ which can be absorbed
renormalizing fields and charge \cite{schw}.

In Eq.(\ref{e1}) for $E>B$ the integral has poles in the $s$-axis which give
origin to an imaginary part corresponding to pair creation. We will be
instead interested in the case of $E=0$ and constant magnetic field for
which
\begin{equation}
\rho =-\int_{0}^{\infty }\frac{ds}{s}\frac{e^{-m^{2}s}}{\left( 4\pi s\right)
^{2}}\left[ \frac{esB}{\sinh \left( esB\right) }-1\right]
\end{equation}
that has no poles.

For the spin-1 $W_{\mu }$ case we adopt the view that the only modification
to $\ln Z$ is the interaction of spin with magnetic field $2e\mathbf{B}\cdot
\mathbf{s}$ in the exponent of Eq.(\ref{e2}). Now the trace must be
performed on momentum and spin states where the latter involves this added
interaction to give
\begin{equation}
\ln Z=\int_{0}^{\infty }\frac{ds}{s}\left( e^{-2eBs}+e^{2eBs}+1\right)
tre^{-\left( -D_{E}^{2}+m_{W}^{2}\right) s}\quad .
\end{equation}
Since the remaining trace is equal to the scalar case, the relevant part of
the vacuum energy density is
\begin{equation}
\rho =-\int_{0}^{\infty }\frac{ds}{s}\frac{e^{-m_{W}^{2}s}}{\left( 4\pi
s\right) ^{2}}\left[ \frac{esB}{\sinh \left( esB\right) }2\cosh \left(
2eBs\right) \right] \quad .
\end{equation}
This expression has no poles but diverges for $s\rightarrow \infty $ when $
B>B_{c}=m_{W}^{2}/e$ due to the gyromagnetic factor $2$ of the $W$ boson,
which would not occur either for $g=1$ or for the $s=1/2$ case. This
divergence is an indication of the vacuum instability for large magnetic
field. The decay rate should be evaluated in the more realistic situation of
$B$ increasing with time, with the consequent generation of an electric
field.

In our calculation of the next section we will not take into account the
evolution with time of the magnetic field but we will consider that when its
value is larger than the critical one in the region of a bubble containing
the broken-symmetry vacuum, the bubble will be destroyed.

\section{Delay of the phase transition due to magnetic field}

It is known that the phase transition of the standard model is of first
order only for Higgs masses which are below the experimental bound \cite{kaj}
. However we will not attempt to use an extension as the MSSM to allow
consistency with this bound since our goal is to establish the effect of a
strong magnetic field on the first-order transition with a reasonably simple
effective potential. In the minimal standard model this temperature
dependent potential for the Higgs field $\varphi $ is \cite{dine}
\begin{equation}
V\left( \varphi ,T\right) =D\left( T^{2}-T_{0}^{2}\right) |\varphi
|^{2}-ET|\varphi |^{3}+\frac{\lambda _{T}}{4}|\varphi |^{4}  \label{e3}
\end{equation}
where $D=\frac{1}{8v^{2}}\left( 2m_{W}^{2}+m_{Z}^{2}+2m_{t}^{2}\right) \quad
,\quad E=\frac{1}{6\pi v^{2}}\left( 2m_{W}^{3}+m_{Z}^{3}\right) \quad ,\quad
$ \newline
$\lambda _{T}=\lambda -\frac{3}{16\pi ^{2}v^{4}}\left( 2m_{W}^{4}\ln \frac{
m_{W}^{2}}{a_{B}T^{2}}+m_{Z}^{4}\ln \frac{m_{Z}^{2}}{a_{B}T^{2}}
-4m_{t}^{2}\ln \frac{m_{t}^{2}}{a_{F}T^{2}}\right) \quad ,\quad $ $T_{0}^{2}=
\frac{1}{4D}\left( m_{H}^{2}-\frac{8B}{v^{2}}\right) \quad ,\newline
B=\frac{3}{64\pi ^{2}}\left( 2m_{W}^{4}+m_{Z}^{4}-4m_{t}^{4}\right) \quad
,\quad \quad $ $\ln a_{B}=3.91\quad ,\quad \ln a_{F}=1.14$.

\strut

All the parameters of the potential Eq.(\ref{e3}) are determined by the
experimental masses of gauge bosons and top quark, $m_{W}=80GeV$, $
m_{Z}=91GeV$, $m_{t}=175GeV$ and the breaking scale $v=246GeV$, with the
only unknown given by the Higgs mass $m_{H}$.

The first-order transition will occur in the range from $T_{c}=\left[ \frac{
T_{0}^{2}}{1-E^{2}/\left( \lambda _{T_{c}}D\right) }\right] ^{1/2}$ to $
T_{0} $, where $T_{c}$ corresponds to equal minima and $T_{0}$ to the
disappearance of the barrier between them. In this range of temperatures
bubbles of broken-symmetry phase are formed with a probability per unit
volume and time given by the semiclassical approximation
\begin{equation}
P\left( T\right) \simeq \omega _{f}^{4}e^{-E_{c}/T}\quad ,
\end{equation}
where $\omega _{f}^{2}=\frac{\partial ^{2}}{\partial \varphi ^{2}}V\left(
\varphi ,T\right) |_{\varphi =0}\,$. $E_{c}$ will come from the maximization
of the bubble energy which, in the thin wall approximation of width $
\varepsilon ,$ is
\begin{equation}
E=-\frac{4\pi }{3}R^{3}\Delta V+4\pi R^{2}\gamma \quad ,  \label{e4}
\end{equation}
where the surface tension is
\begin{equation}
\gamma \simeq \frac{\left( \Delta \varphi \right) ^{2}}{\varepsilon }\quad .
\end{equation}
$\Delta V$ and $\Delta \varphi $ are the differences of potential and Higgs
field between the two phases respectively. The maximization of Eq.(\ref{e4})
gives the critical radius $R_{c}$ and energy $E_{c}$
\begin{equation}
R_{c}=\frac{2\gamma }{\Delta V}\quad ,\quad E_{c}=\frac{16\pi }{3}\frac{
\gamma ^{3}}{\left( \Delta V\right) ^{2}}\quad .
\end{equation}
The velocity of expansion of the bubble can be estimated by \cite{linde}
\begin{equation}
v_{W}\left( T\right) \simeq \frac{\Delta V}{T^{4}}\quad ,
\end{equation}
and the wall width by $\varepsilon \simeq m_{H}\left( T\right) ^{-1}$ where $
m_{H}\left( T_{c}\right) \simeq m_{H}/50$, for not too high values of $m_{H}$
.

We will consider that the phase transition will be completed when the
average distance between centres of bubbles $2R_{n}\left( t\right) $ is
equal to twice their average radius $R_{b}(t)$, i.e. when \newline
$d\left( t\right) =2\left[ R_{n }\left( t\right) -R_{b}\left( t\right)
\right] $ vanishes (Alternatively, one could determine the moment at which
the fraction of volume occupied by bubbles of new phase equals one \cite
{linde, guth}). The relation between the time $t$ and temperature in the
radiation regime is taken as
\begin{equation}
t=k/T^{2}\quad ,\quad k\sim 10^{16}GeV\quad ,
\end{equation}
to give the beginning of the phase transition at $t\sim 10^{-12} sec $.

In terms of the probability for bubble formation, its number density will be
\begin{equation}
n \left( t\right) =\int_{t_{c}}^{t}dt^{^{\prime }}P\left( t^{^{\prime
}}\right) \quad ,
\end{equation}
so that the average radius of the volume per bubble is
\begin{equation}
R_n\left( T\right) =\left[ \frac{4}{3}\pi n \left( T\right) \right]
^{-1/3}\quad .
\end{equation}
The average bubble radius comes from
\begin{equation}
R_{b}\left( t\right) =\frac{1}{n \left( t\right) }\int_{t_{c}}^{t}R\left(
t^{^{\prime }},t\right) P\left( t^{^{\prime }}\right) dt^{^{\prime }}\quad ,
\label{rb}
\end{equation}
where $R\left( t^{^{\prime }},t\right) =R_{c}\left( t^{^{\prime }}\right)
+v_{W}\left( t^{^{\prime }}\right) \left( t-t^{^{\prime }}\right) $.

\strut

Now we must include the influence of the magnetic field generated during the
expansion of bubbles. Several mechanisms of creation of seed fields (e.g.
charge separation in bubble walls, bubble collisions or fluctuating Higgs
gradients) \cite{baym, cheng} involve the appearance of non-zero expectation
values of the Higgs field. This means that the seed fields will arise in
general in the regions where bubbles are present. Once a seed is formed, the
magnetic field is quickly amplified due to magnetic turbulence in the
conducting fluid, until it reaches in this region the value given by energy
equipartition. Therefore we will consider that when a bubble is produced a
magnetic field
\begin{equation}
B\simeq c\frac{T^{2}}{\left( vR\right) ^{1/2}}
\end{equation}
will manifest itself in a region $R$ after a time $\tau $. The exponent of
the denominator is the statistically most favorable one to have sizeable
homogeneous fields as the region increases. The constant $c$ has a maximum
value of $100$ to satisfy the constraint given by nucleosynthesis \cite
{grass}. We will assume $\tau \simeq R_{c}/v_{W}$ which is in reasonable
agreement with some mechanisms of magnetic field generation \cite{baym}.
Since for $B\gsim B_{c}$ the decay rate of the electroweak vacuum, $\Gamma
\sim (eB_{c})^{2}$ \cite{amb}, is much larger than the bubble nucleation or
expansion rates, we can consider that the bubble disappears because the
symmetry inside it is restored. In this way $R(t^{\prime },t)=0$ for $
t^{^{\prime }}+\tau <t$ and $B>B_{c}=m_{W}^{2}/e$ in (\ref{rb}) and the
completion of the phase transition according to a bubble mechanism will be
either delayed or avoided.

Our numerical calculation shows that for large values of $m_{H}$, e.g. $
200GeV$, the first-order phase transition is not completed even without
magnetic field (see Fig.1).

For a low value of the Higgs mass $m_{H}\simeq 70GeV$ our model allows the
completion of the first-order transition without magnetic field (see Fig.2).
If the maximum possible value of the magnetic field $c=100$ is assumed, the
first-order transition is avoided (Fig.2a). With a smaller value of the
magnetic field, e.g. $c=36$, the completion of the first-order transition is
only delayed in around $100MeV$ (Fig.2b), and well below this intensity the
effect of the magnetic field is negligible.

An alternative scenario would be to consider that the magnetic field was
generated in a prior stage of the universe, e.g. the GUT epoch. Its effect
on the electroweak phase transition can be studied through an effective
potential depending on temperature and constant magnetic field \cite{kainu}.
It is interesting that again for $m_{H}\lsim80GeV$ the magnetic field delays
the first-order transition due to an increase of the free energy of the
broken-symmetry phase. This result agrees with the tendency shown by our
calculation but we remark that the two situations are different. In our case
we do not have a magnetic field prior to the phase transition but a strong
field is formed in small regions around the bubbles, destroying them. The
magnetic field in a subhorizon volume will have a statistical average
several orders of magnitude below the critical value, therefore its
influence on the effective potencial will be negligible. If the first-order
transition is not completed, at temperature $T_{0}$ our system will be at
the maximum of the potential $\varphi \sim 0$ whereas the minimum will be at
$\varphi \neq 0$ without any barrier. Now the phase transition will occur
through the global rolling down of the field $\varphi $ towards the minimum
of the potential.

\begin{figure}[tbp]
\centering
\epsfxsize=9cm \epsfbox{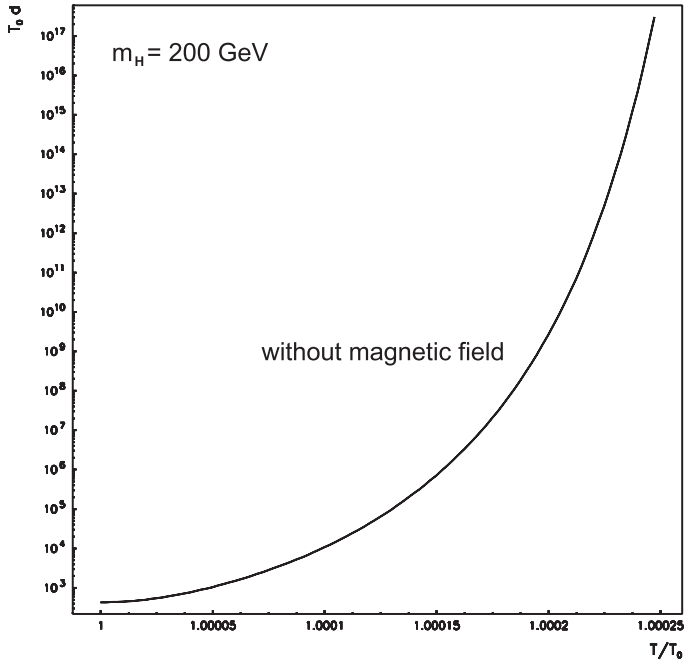}
\caption{Mean distance between the walls of two bubbles in units of $1/T_{0}$
versus temperature in units of $T_{0}$ for $m_{H}=200GeV$.}
\end{figure}

\begin{figure}[tbp]
\epsfxsize=17cm \epsfbox{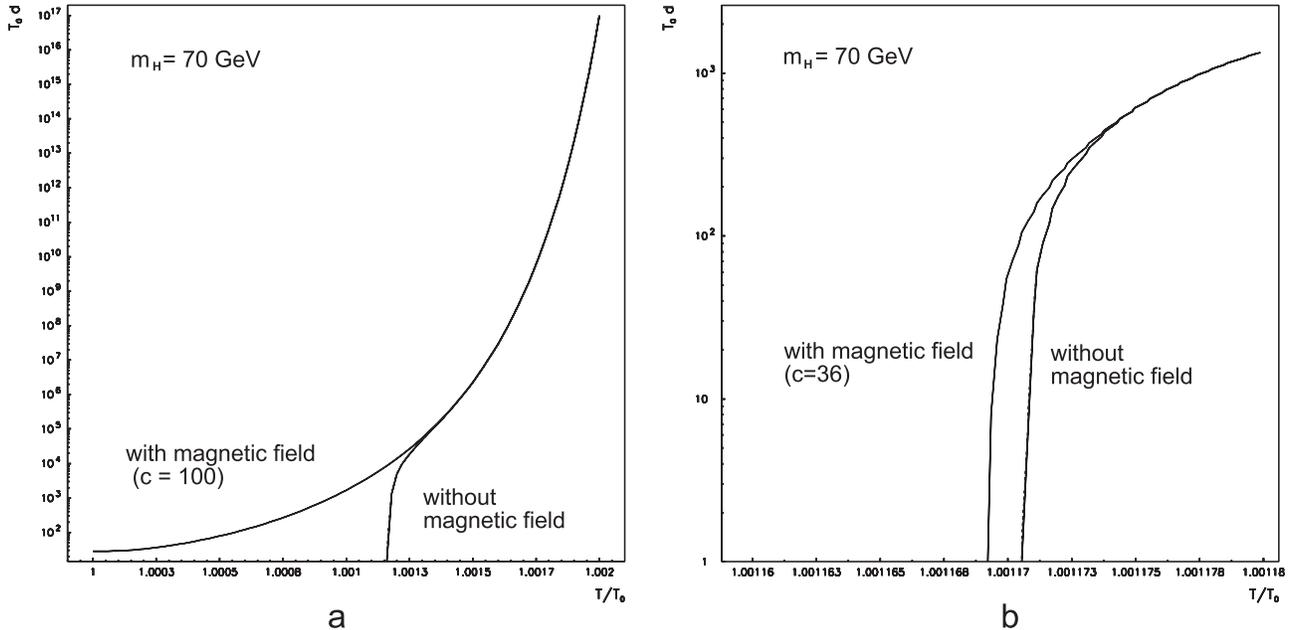}
\caption{Mean distance between the walls of two bubbles in units of $1/T_{0}$
versus temperature in units of $T_{0}$ for $m_{H}=200GeV$. \textbf{a.} High
magnetic field ($c=100$). \textbf{b.} Low magnetic field ($c=36$).}
\end{figure}

\section{Conclusions}

We have seen that the highest possible magnetic field together with the most
favorable law for having homogeneous field in regions of increasing size
might have cosmological consequences through the non-completion of the
first-order electroweak transition through a bubble mechanism. Therefore the
usual electroweak baryogenesis due to bubble expansion would be affected.
But one must notice that also the homogeneous increase of $\varphi $ can
produce a matter-antimatter asymmetry. This is because there will be a
baryonic chemical potential related to the time variation of the $CP$
violating phase $\theta $. The resulting baryonic density will depend on the
variation $\Delta \theta $ in the interval when the sphalerons are active
due to the smallness of $\varphi $. For a weakly first-order transition, an
advantage of this mechanism compared to the bubble one is that the baryonic
density would not be erased in the broken phase because here the value of $
\varphi $ is larger due to the delay of the phase transition. One may remind
that this problem is also avoided by the baryogenesis in cosmic strings but
paying the price of a suppression factor in the active volume.

However, it is unlikely that such a strong and large size primordial
magnetic field has occurred, and for more acceptable fields the effect would
be only a small decrease of the temperature for the completion of a
first-order transition.

We have studied the influence of the magnetic field on the phase transition
using the easiest model, i.e. the standard model and not the MSSM where
presumably the first-order phase transition can occur for not too light
Higgs mass \cite{care}. The fact that we obtain the completion of the
first-order transition without magnetic field for $m_{H}\simeq 70GeV$, not
far below the experimental bound is probably due to the definition that it
occurs when the bubbles touch each other without taking into account their
scattering. But we believe that the general
conclusions on the magnitude of the effect do not depend on the details of
the used electroweak model.

Regarding further developments of this calculation, it would be important to
evaluate the vacuum decay rate caused by a time dependent magnetic field in
order to consider more carefully its effect on the bubbles instead of taking
the simplification of assuming their disappearance as soon as the magnetic
field is larger than the critical value.

\strut

\strut

\strut

\textbf{Acknowledgments}

We are deeply indebted to A. Della Selva and R. Iengo for very stimulating
discussions. L. M. and A. M. thank the kind hospitality of the Departments
of Physics of the Universities of Napoli and della Calabria at Cosenza where
parts of this research has been performed.

This work was partially supported by CONICET through PICT 0358.

\end{document}